\documentclass[final,5p,times,twocolumn]{elsarticle}

\usepackage{amssymb,amsbsy,amsmath}

\begin{document}

\begin{frontmatter}

\title{Molecular dynamics and continuum analyses of the electrokinetic zeta potential\\ in nanostructured slit channels}

%% Group authors per affiliation:
\author[mainaddress]{Sijia Huagn\fnref{fn1}}
\author[mainaddress]{Amir M. Rahmani\fnref{fn1}}
\author[mainaddress]{Troy Singletary}
\address[mainaddress]{Department of Mechanical Engineering, Stony Brook University, Stony Brook, NY 11794, USA}
\address[secondaddress]{Department of Applied Mathematics \& Statistics, Stony Brook University, Stony Brook, NY 11794, USA}
\fntext[fn1]{The first two authors contributed equally to this work.}

\author[mainaddress,secondaddress]{Carlos E. Colosqui\corref{mycorrespondingauthor}}
\cortext[mycorrespondingauthor]{Corresponding author}
\ead{carlos.colosqui@stonybrook.edu}

\begin{abstract}
This work presents a theoretical and numerical study of electrokinetic flow and the zeta potential for the case of slit channels with nanoscale roughness of dimensions comparable to the Debye length, by employing molecular dynamics simulations and continuum-level analyses.
A simple analytical model for considering the effects of the surface roughness is proposed by employing matched asymptotic solutions for the charge density and fluid flow field, and matching conditions that satisfy electroneutrality and the Onsager reciprocal relation between the electroosmotic flow rate and streaming current. 
The proposed analytical model quantitatively accounts for results from molecular dynamics simulations that consider the presence of ion solvation shells and surface hydration layers.
Our analysis indicates that a simultaneous knowledge of the electroosmotic and pressure-driven flow rate or streaming current can be instrumental to determine unambiguously the zeta potential and the characteristic surface roughness height.
\end{abstract}

\begin{keyword}
\texttt{Electrokinetic flow}\sep zeta potential \sep molecular dynamics
\end{keyword}

\end{frontmatter}

%\linenumbers

\section{Introduction}

Electrokinetic phenomena are due to the coupling of fluid and charge transport in the Electric Double Layer (EDL) that forms at the interface between a charged surface and a liquid electrolyte solution.\cite{levich1962,probstein2005}
These phenomena combining electrodynamic and hydrodynamic effects enable ground-breaking technologies for separation processes and actuation, such as micellar electrokinetic chromatography,\cite{quirino1998,hompesch2005analysis,silva2013micellar}  electrophoretic assembly of bio/nanomaterials,\cite{hayward2000,velev2006c,siavoshi2011size} and electrodialytic water desalination,\cite{lee2002designing,sadrzadeh2008,almarzooqi2014application}. 
Electroosmostic flow and electrophoresis are prototypical examples of electrokinetic phenomena where applied electric fields induce the controlled motion of fluid and/or colloidal solutes (e.g., nanoparticles, macromolecules). 
At the same time, applied pressure gradients produce the flow of electrolyte or colloidal solutions and the advection of ion (or colloidal solutes) produce so-called streaming or (sedimentation) currents.

Both the fluid flow driven by voltage and electric current driven by pressure differences are determined by the so-called zeta potential $\zeta$, which is a metric of the electrokinetic coupling in the EDL that is measurable by different experimental methods.\cite{elimelech1994measuring,sze2003zeta,delgado2007}
Conventional theoretical descriptions for electrokinetic flows, and thus analytical models for the zeta potential, have been developed based on the classical continuum assumptions of sharp liquid-solid interfaces of plane or spherical surfaces that are physically smooth and chemically homogeneous.
As a result, and despite insightful theoretical and computational studies, \cite{cummings2000conditions,hu2003electrokinetic,qiao2007effects,yang2008numerical,liu2010molecular,messinger2010,yoshida2016a} there are no well-established analytical models to account for experimental observations and/or predict analytically the zeta potential for surfaces with nanoscale physical structures or roughness of arbitrary shape and dimensions comparable to the Debye length $\lambda_D$. 

For the particular case of electrokinetic flow and zeta potentials in slit channels, analytical models are relatively well-developed and, to a certain extent, validated against experimental observations for the case of flat surfaces with uniform surface charge density.\cite{kirby2004,kirby2004b}  
For the case of surfaces with roughness of characteristic dimension $\delta$, it has been theoretically established that when $\delta\ll\lambda_D$ the surface roughness has no effect on the zeta potential in the case of negligible surface conductivity \cite{cummings2000conditions} and suppresses the zeta potential for the case of finite surface conductivity.\cite{messinger2010} 
Recent experimental work on nanostructured surfaces by our group corroborates this theoretical findings.\cite{aktar2019} 
Nevertheless, no analytical expression is available to predict the zeta potential for the case of finite surface conductivity and surface nanostructure or roughness dimensions comparable to the Debye length. 

This work proposes the application of a simple analytical description for streamwise-averaged unidirectional electrokinetic flow that is valid for the case of slit channels with periodic nanoscale roughness and heterogeneous surface charge of characteristic height comparable to the Debye length.
The key element of the proposed analytical model is the use of matched asymptotic solutions and the enforcement of charge and volume conservation, and the Onsager reciprocal relation in electrokinetic flows, through the matching conditions.
Analytical estimates from the proposed model for the streamwise-averaged charge density and flow velocity profiles, and the zeta potential are in close agreement with results from non-equilibrium molecular dynamics (MD) simulations of nanoscale channels with a modeled sinusoidal roughness of period 
$\ell >\lambda_D$ and height $\delta \gtrsim \lambda_D$.  
Our theoretical analysis and computational results indicate that the zeta potential and characteristic surface roughness height can be unambiguously determined, provided that the Onsager reciprocal relation between electroosmotic flow and streaming current is satisfied, and a reliable estimate for the pressure-driven flow rate is available.

\section{Continuum level analysis \label{sec:continuum}}
\begin{figure}[h]
\centering
  \includegraphics[width=0.48\textwidth]{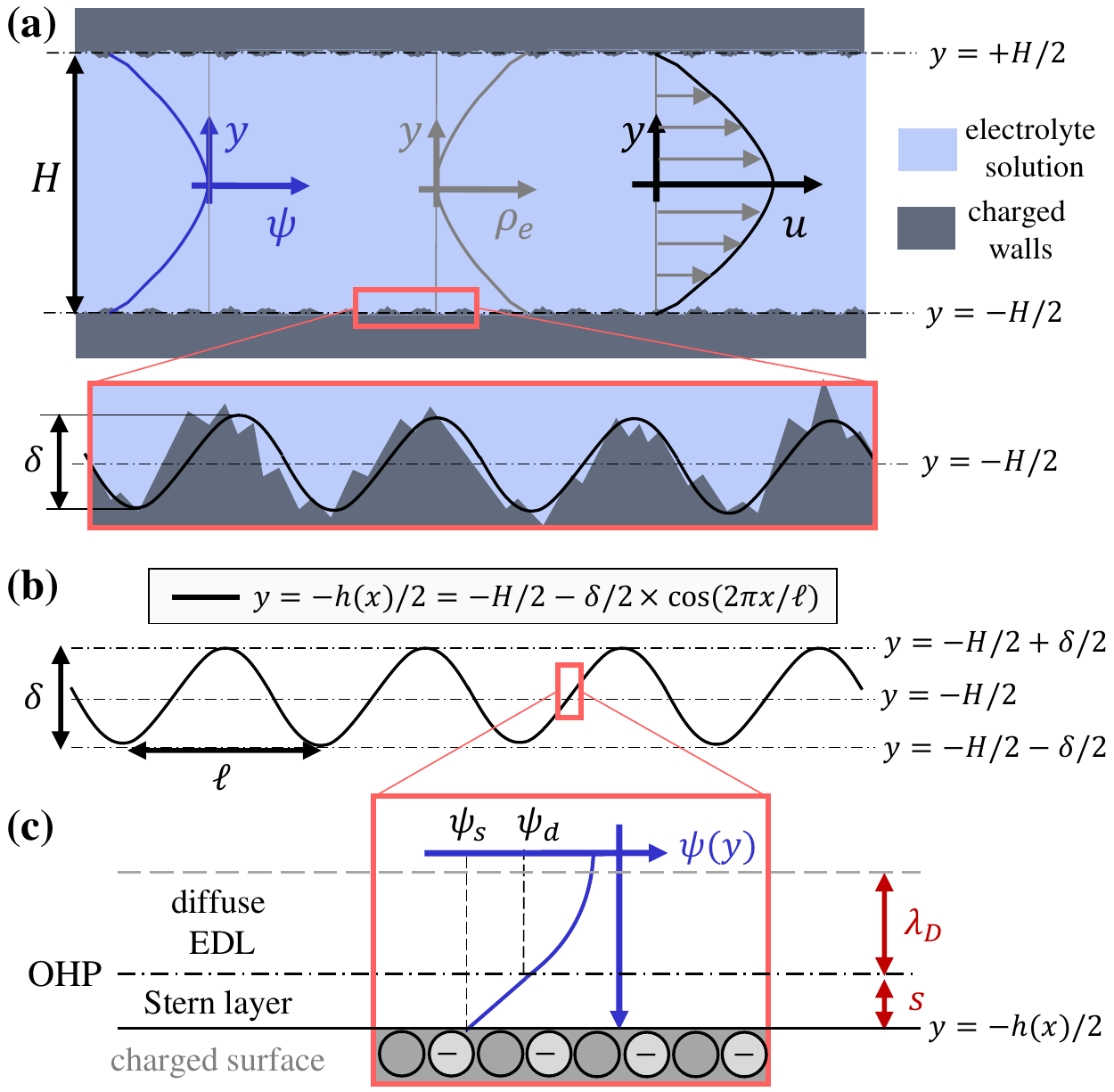}
\caption{
Electrokinetic flow in slit nanochannels with nanoscale surface roughness.
(a) Effective or streamwise-averaged profiles for the electrostatic potential $\psi(y)$, charge density $\rho_e(y)$, and flow velocity $u(y)$.
The channel has a nominal or average height $H$ and nanoscale surface roughness of characteristic height $\delta$.
(b) The effective surface roughness is modeled as a zero-mean single-mode perturbation on the local channel height $h(x)$.  
(c) Illustration of the Electric Double Layer structure at each point of the charged solid surface. 
The Outer Helmholtz Plane (OHP) separates the diffuse and Stern layer. 
}
  \label{fig:1}
\end{figure}

Within the framework of continuum analysis, we formulate a simple and compact description for electrokinetic flows that can accurately predict the zeta potential for the system illustrated in Fig.~\ref{fig:1}a, which consists of a 3D slit nanochannel of nominal or average height $H$, length $L\gg H$, and width $W\gg H$, with both channel surfaces having a nanoscale regular structure and/or random roughness having a characteristic height $\delta\lesssim H$. 
The proposed description is based on streamwise-averaged variables (i.e., the effective ion density and flow velocity profile) that are obtained by integration along the $x$- and $z$-coordinates and can predict accurately the total number of ions and the volumetric flow rate when integrated along the channel height.

The entire channel is filled with a liquid solution of a 1:1 symmetric electrolyte, having molar concentration $c$, shear viscosity $\mu$, and permittivity $\varepsilon$ in the bulk. 
We analyze only the case that the channel surfaces are highly wettable and therefore the full surface area is in direct contact with liquid; i.e., both channel walls are in the Wenzel wetting state \cite{wenzel1936resistance,ishino2004wetting}, as illustrated in Fig.~\ref{fig:1}a.
For the sake of analytical simplicity, the nanoscale surface structure or roughness is approximately modeled by a single-mode and zero-mean perturbation (Fig.~\ref{fig:1}b) with a characteristic amplitude $\delta$ and a period or coherence length $\ell$ that can be obtained by averaging along the transverse $z$-direction and/or filtering with a proper cut-off wavelength the surface topography height.
For the sake of analytical simplicity, we thus consider that the effective local height of the nanochannel varies according to $h(x)=H+\delta \cos(2\pi x/\ell)$.

In the present analysis it is assumed that each channel wall has a known number $N_w$ of (negatively) charged sites (Fig.~\ref{fig:1}c) distributed with a uniform surface charge density $\sigma=-e N_w/W L$ ($e$ is the elementary charge), which circumvents the need for using a site-dissociation model relating $\sigma$ with the surface potential $\psi_S$.\cite{healy1978,hiemstra1996intrinsic,borkovec1997origin,behrens2001charge}  
Following the Stern-Gouy-Chapman (SGC) model,\cite{chapman1913,stern1924} the EDL formed at any point of the charged solid-liquid interface (Fig.~\ref{fig:1}c) comprises (i) the Stern layer with immobile liquid molecules and ions having an effective diameter $s$, and (ii) the diffuse layer having mobile ions in solution and a characteristic thickness determined by the Debye length $\lambda_D$.
Assuming near equilibrium thermodynamic conditions and a system temperature $T$, the local number density of anions $n_{-}$ and cations $n_{+}$ is approximately given by Boltzmann statistics,
$n_{\pm}=n_0 
\exp\left(\mp\frac{e\psi}{k_B T}\right)
\exp\left(-\frac{U_{S}}{k_B T}\right), 
$
where $n_0$ is the zero-potential number density of anions and cations, $\psi$ is the electrostatic potential, and 
$U_S$ is the free energy due to ion solvation, surface hydration, and any other non-electrostatic interactions ($k_B$ is the Boltzmann constant).  
The number density of solvent molecules $n_S$ presents large spatial oscillations with respect to the bulk liquid density $n_{0S}$ at distances from the wall of a few molecular diameters $s$ due to the presence of non-electrostatic interactions associated with the formation of solvation shells and hydration layers results.\cite{pashley1982hydration,israelachvili1983molecular,israelachvili2015}

In the SGC description of the EDL, it is assumed that the co-ions closest to the wall are located in the Outer Helmholtz Plane (OHP) separating the Stern and diffuse layer (Fig.~\ref{fig:1}c) and no significant number of ions are present within the Stern layer where $U_S\gg k_B T$ as a result of strong solvation and hydration interactions.  
Accordingly, the electrostatic potential $\psi$ in the Stern layer is assumed to decay linearly with the distance from the surface and the diffuse layer potential 
$\psi_d=\psi_S-\sigma/C_S$ at the OHP (Fig.~\ref{fig:1}c) is prescribed by the specific Stern capacitance $C_S=\varepsilon_S/s$, where $\varepsilon_S$ is the permittivity in the Stern layer.
Outside the Stern layer where the medium permittivity is $\varepsilon\simeq$~const. and $|\psi|>>U_S$, the local charge density is $\rho_e=e(n_{+}-n_{-})=-2 e n_0 \sinh(e\psi/k_B T)$ and the electrostatic potential satisfies the Poisson-Boltzmann (P-B) equation 
$\varepsilon \nabla^2 \psi=2 e n_0 \sinh(e\psi/k_B T)$.\cite{levich1962,probstein2005}

In the conventional continuum description for steady-state and incompressible electrokinetic flow the P-B equation supplements the Navier-Stokes (N-S) momentum equation, 
$\mu \nabla^2 {\bf u} -\rho_e \boldsymbol\nabla \psi - \boldsymbol\nabla p=0$.
However, the assumption of a constant shear viscosity $\mu$ employed in the N-S is strictly valid within the fluid bulk outside the hydration layer where liquid molecules are strongly adsorbed and become immobilized at the solid-liquid interface. 
Considering that the effective viscosity becomes extremely large and thus $\nabla^2 {\bf u}\simeq 0$ within the Stern layer, we will assume a linear flow velocity profile at sufficiently small distances from the wall.
%

%Despite the effectiveness of this modeling assumption, solvation and hydration interactions produce strong spatial oscillations of the solvation energy $U_S$ within distances from the liquid-solid interface that comparable to the solvent molecule diameter $s$,\cite{solvation_oscillation} which is indeed observed in the ion density profiles reported by MD simulations (e.g., see Results and discussion).
%

\subsection{Electrokinetic flow equations} 
In this section, we formulate analytical expressions for the streamwise-averaged or ``effective'' electrostatic potential $\psi(y)$, ion number density $n_\pm(y)$, unidirectional velocity $u_P(y)$ for pressure-driven (Poiseuille) flow, and $u_E(y)$ for voltage-driven (electroosmotic) flow in a slit channel with a modeled single-mode variation of the effective height $h(x)$ that is induced by nanoscale surface roughness of amplitude $\delta$ and period $\ell$.  
We assume that nanoscale roughness of height $\delta \lesssim 10 s$ produces similar effects as the hydration layer in immobilizing solvent molecules and preventing adsorption of counterions at the liquid-solid interface; the validity of this assumption will be assessed by comparison against results from MD simulations.
Hence, the effective ion density and flow velocity profiles is modeled as linear functions decaying with the distance from the solid-liquid interface for which $|y|\gtrsim H/2-\delta/2$
Approximate solution of the P-B and N-S equation is thus obtained in the form of matched asymptotic expressions valid for $0\le |y|\le y^*$ (i.e., the ``outer'' region) and for $y^*\le|y|\le H^*$ (i.e., ``inner'' region), where $y^*=H/2-\delta/2-s/2$ and $H^*=H/2+\delta/2$.
The coordinate $y^*$ of the mathching point is thus located half a molecular diameter away from the cuspid of the surface defect of height $\delta$ due to the presence of the Stern layer, which comprises the first hydration layer with immobile solvent atoms.

The proposed analytical expressions for streamwise-averaged variables in the outer region are obtained via asymptotic analysis for $\ell\gg\lambda_D$ and their deviations from the 3D physical variables are of order ${\cal O}(\lambda_D/\ell)$.
Analytical solution of the nonlinear P-B equation in the outer region gives the electrostatic potential profile\cite{behrens1999}
\begin{equation}
\psi(y)=\psi(0)+2 \ln \left[ \mathrm{cd}\left( \frac{y}{2\lambda_D} e^{\frac{-e\psi(0)}{2 k_BT}},e^{\frac{2e\psi(0)}{k_BT}} \right) \right],
\label{eq:psi}
\end{equation}
where $\psi(0)$ is the centerline potential at $y=0$, $\lambda_D=\sqrt{\varepsilon k_B T/2 e^2 n_0}$ is the Debye length, and $\mathrm{cd}(a,m)$ is the cd Jacobi elliptic function with argument $a$ and parameter $m$.
The streamwise-averaged ion density is given by
\begin{equation}
n_\pm(y)=\begin{cases}
n_0 \times \exp\left(\mp\frac{e \psi}{k_B T}\right) 
&\!\! \mathrm{for}~0\le\!|y|\!\le\!y^*\\[10 pt]
n_0 \times \exp\left(\mp\frac{e \psi^*}{k_B T}\right) \times \frac{(|y|-y^*)}{\delta} 
&\!\! \mathrm{for}~y^*\!\le\!|y|\!\le H^*
\end{cases}
\label{eq:ions}
\end{equation}
where $\psi^*=\psi(y^*)$ is the electrostatic potential at the boundary between the inner and outer region.
When a pressure head $\Delta p$ and/or voltage difference $\Delta V$ are applied across the channel, the resulting streamwise-averaged flow velocity is 
\begin{equation}
u(y) =\begin{cases}
\frac{\varepsilon \Delta V}{\mu L}\! \left[\psi(y)-\psi^*\right]
+\frac{\Delta P}{2\mu L}\!\left[\frac{H^2}{4}-y^2\right]+u^*
&\!\!\!\!\!\! \mathrm{for}~0\le\!|y|\!\le\!y^*\\[10 pt]
u^* \times \frac{(|y|-y^*)}{\delta} 
&\!\!\!\!\!\! \mathrm{for}~y^*\!\le\!|y|\!\le\!H^*
\end{cases}
\label{eq:flow}
\end{equation}
where $u^*=u(y^*)$ is the flow velocity at the matching coordinate.

The description in Eqs.~\ref{eq:psi}--\ref{eq:flow} can be employed to predict the total number of (positive/negative) charges in solution
\begin{equation}
N_\pm=2WL\int_{0}^{y^*} n_\pm(y) dy+WL\times n_\pm^*\times\delta,
\label{eq:charges}
\end{equation}
and the volumetric flow rate
\begin{equation}
Q(\Delta p,\Delta V)=2W\int_{0}^{y^*} [u(y)-u^*] dy+WH\times u^*.
\label{eq:Q}
\end{equation}
Furthermore, the streaming current can be estimated from
\begin{equation}
I_s=2W\int_{0}^{y^*} \rho_e u_P(y) dy+\frac{2W}{3}\times\rho_e^*u_P^*\times\delta,
\label{eq:Is}
\end{equation}
where $u_P=u(\Delta p, \Delta V=0)$ is the velocity profile in Eq.~\ref{eq:flow} for pressure-driven flow, $u_P^*=u_P(y^*)$ and $\rho_e^*=\rho_e(y^*)$.

\subsection{Determination of the zeta potential}

The integration of streamwise-averaged variables in Eqs.~\ref{eq:charges}--\ref{eq:Is} can be performed to estimate the electrokinetic zeta potential for the studied slit nanochannel (Fig.~\ref{fig:1}). 
Under conditions for which the Onsager reciprocal relations are valid,\cite{mazur2013,brunet2004generalized} the zeta potential is equally defined by the electroosmotic flow rate $Q_e=Q(0,\Delta V)$ driven by voltage and streaming current $I_s(\Delta p)$ driven by a pressure, according to
\begin{equation}
\zeta
=-\frac{\mu}{\varepsilon}\times\frac{Q_e}{W H}
\times\frac{L}{\Delta V}
=-\frac{\mu}{\varepsilon}\times\frac{I_s}{W H}
\times\frac{L}{\Delta p}.
\label{eq:zeta_equality}
\end{equation}

In order to produce quantitative predictions for the zeta potential via Eq.~\ref{eq:zeta_equality} it is necessary to determine the zero-potential ion density $n_0$, the centerline potential $\psi_0$, and the fluid flow velocity $u^*$ at the matching coordinate $y^*$.
These three parameters can be readily obtained by solving Eqs.~\ref{eq:charges}--\ref{eq:Q} when $N_{+}$, $N_{-}$, and $Q$ or $Q_e$ are known, and an estimate for the effective roughness height $\delta$ is available (e.g., from topographic data analysis). 
Moreover, it is possible to obtain estimates for the effective roughness $\delta$ from Eq.~\ref{eq:Is} and the reciprocal relation $Q_e/\Delta V=I_s/\Delta p$, when the pressure-driven flow rate $Q_P=Q(\Delta p,0)$, and thus $u_P^*$, are known.
The proposed expressions and analysis for slit channels with nanoscale roughness will be employed to account for results from MD simulations described in the next section.

\section{Molecular dynamics simulations \label{sec:MD}}
\begin{figure*}[t]
\centering
  \includegraphics[width=1\textwidth]{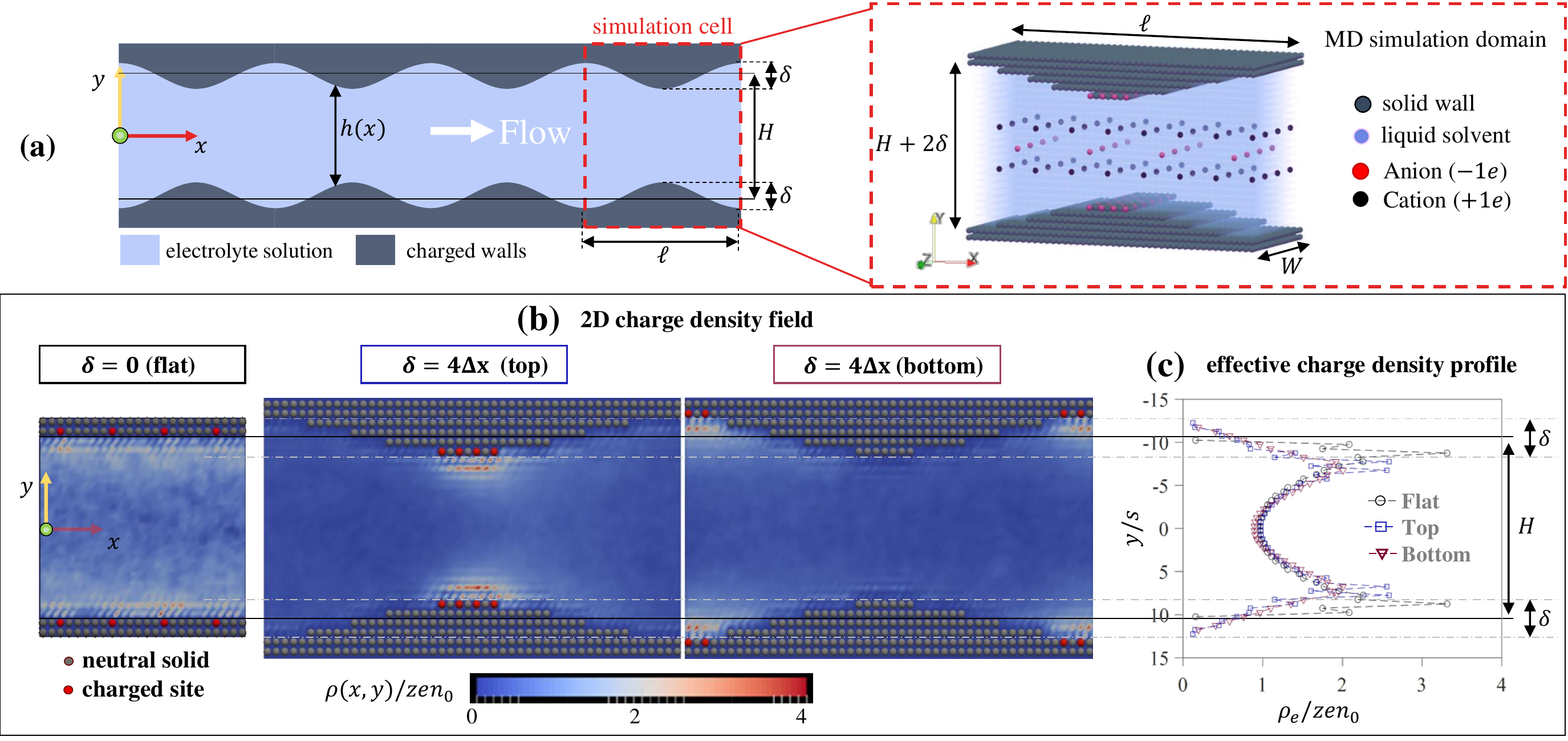}
  \caption{Molecular Dynamics (MD) simulation setup.
(a) The simulation domain is a section of a slit nanochannel of average height $H$ with with periodic surface features of height $\delta$ and period $\ell$.
The simulation comprises for different atomic species corresponding to the solid wall, liquid solvent, anions, and cations. 
(b) Two-dimensional charge density field $\rho(x,y)$ for the modeled ``flat'' and ``rough'' channel configurations.
The surface charge is modeled by anions fixed to the wall in different configurations, with charges concentrated at the cuspid or bottom of the modeled surface features. 
(c) Effective of streamwise-averaged charge density profile $\rho_e(y)$. 
Under the studied conditions with $\ell>\lambda_D$ different surface charge distributions (top/bottom) produce a similar effective charge density profile.    
}
  \label{fig:2}
\end{figure*}

In this work, non-equilibrium MD simulations in the NVT ensemble are performed with the open-source package LAMMPS\cite{plimpton1993fast,plimpton2007lammps,lammps} in order to model electrokinetic flow in a 3D section of a nanoscale slit channel (see Fig.~\ref{fig:2}a) of average height $H=20 \Delta x$ and width $W=10 \Delta x$, with surfaces that are either flat ($\delta=0$) or have periodic sinusoidal features of height $\delta= 4\Delta x$ and period $\ell=48 \Delta x$. 
In all cases, periodic boundary conditions are applied in the $x$- and $z$-directions, the lattice spacing employed is $\Delta x=s\times 0.8^{-1/3}$ and the molecular diameter is $s=3.5$ \AA. 
The modeled system comprises four atomic species corresponding to the the liquid solvent ($i=1$), cations ($i=2$) with unit charge ($q=+e)$, anions and negatively charged surface sites ($i=3$) with unit charge ($q=-e)$, and solid channel walls ($i=4$); all atomic species have the same atomic mass and diameter $s$. 
The interaction energy between two atoms species $i$ and $j$ separated by a distance $r$ is determined by the sum of pairwise Lennard-Jones (L-J) and Coulomb potentials, 
$u_{ij}(r)=4\epsilon_{ij}[(s/r)^{12}-(s/r)^{6}]-C_eq_iq_j\times (s/r)$,
with the interaction energies $\epsilon_{ij}$ and charges $q_i$ described in Table~\ref{tab:1}, and the characteristic electrostatic energy $C_e=e^2/(4\pi\varepsilon s)=$~1 and 5 $k_B T$, which models two different conditions with a different dielectric permittivity.
For the sake computational efficiency, pairwise L-J interactions are only computed within distance $r/s\le 2.5$ and Coulomb interactions are computed explicitly only within distances $r/s\le 5$, beyond which the electrostatic interaction energy is computed using the Particle-Particle Particle-Mesh (PPPM) method with a relative error set to 10$^{-5}$ for the computed electrostatic force.\cite{hockney1988}  
To simulate pressure-driven flow a force $f_g=0.235$ pN is applied in the positive $x$-direction to all solvent atoms and ions, while electroosmotic flow is produced by a body force of magnitude 
$f_e=11.8$ pN that is applied in the positive $x$-direction to the cations and in the opposite direction to the anions.
The applied forces have the equivalent effect of a pressure head $\Delta p=N f_g/W H$ ($N=22176$ is the total number of solvent atoms and ions) and a voltage difference $\Delta V=f_e L/e$. 
A time step resolution of 1 fs is employed and the total simulation time is 10 million time steps, which corresponds to a physical time of 10 ns.
A linear response between the steady-state volumetric flow rate and the force magnitude is observed under the studied conditions.

\begin{table}[h]
\small
  \caption{\ Lennard-Jones interaction energies $\epsilon_{ij}$ and Coulombic charges $q_i$ employed in MD simulations for each atomic species ($i=1,4$).}
  \label{tbl:example}
  \begin{tabular*}{0.48\textwidth}{@{\extracolsep{\fill}}lcccc}
    \hline
    $\epsilon_{ij}/k_BT$& solvent & cation & anion  & solid \\
    \hline
    solvent   & 1 & 1.4 & 1.4 & 1\\
    cation  ($i=2$) & 1.4 & 1 & 1 & 1\\
    anion ($i=3$) &1.4 & 1 & 1 & 1\\
    solid ($i=4$) & 1 & 1 & 1 & 1\\
    \hline
    charge $q_i/e$ & 0 & +1 & -1 & 0\\
    \hline

  \end{tabular*}
\label{tab:1}
\end{table}

The modeled system comprises $N_-=16$ anions and $N_+=64$ cations in solution, and $N_w=48$ negatively charged sites fixed on the solid walls.
Hence, the system is electroneutral since the total number cations in solution includes those coming from the dissociation of both the electrolyte and the chargeable surface sites.
The studied condition corresponds to a 1:1 electrolyte solution of molar concentration $c\simeq 20$ mM, and the Debye length is $\lambda_D=1.98$ nm for the case with the weakest electrostatic interactions ($C_e=k_B T$) and $\lambda_D=0.9$ nm for the case with the strongest electrostatic interaction ($C_e=5 k_B T$). 
The surface charge density is $\sigma=-23.5$ mC/m$^2$ in all the modeled cases (see Fig.~\ref{fig:2}b), which approximately corresponds to the surface charge of silica at pH~$\simeq$~8.\cite{behrens2001charge,aktar2019}
For the flat channel configuration the negatively charged sites are uniformly distributed, which is consistent with continuum model hypothesis of homogeneously charged surface. 

To assess the effects of inhomogeneity in the distribution of surface charges, MD simulations for the slit channel with sinusoidal surface features were performed with the charged sites localized at either the cuspid or bottom of the roughness (cf. Fig.~\ref{fig:2}b).
Theoretical and experimental analyses\cite{weinstein1953,heinz1999charge,gongadze2011adhesion,drelich2011charge} indicate that, depending on the physicochemical conditions, surface charge formation tends to concentrate on the cuspid of surface features where the convex curvature induces the strengthening of the local surface potential.
For the conditions modeled in this work and in the limit cases of all surface charges at the cuspid or at the bottom of the surface features (Fig.~\ref{fig:2}b), there are no significant differences in the streamwise-averaged profiles of the charge density profile (cf. Fig.~\ref{fig:2}c) and other studied variables, as predicted analytically for $\ell/\lambda_D\ll 1$.
Independently of the localization of the surface charges, the modeled nanoscale roughness has similar noticeable effect on the streamwise-averaged variables (cf. Fig.~\ref{fig:2}c) that is analogous to the formation of a region of approximate thickness $\delta$ with a nearly linear profile. 

\section{Results and discussion}
\begin{figure}[t!]
\centering
  \includegraphics[width=0.48\textwidth]{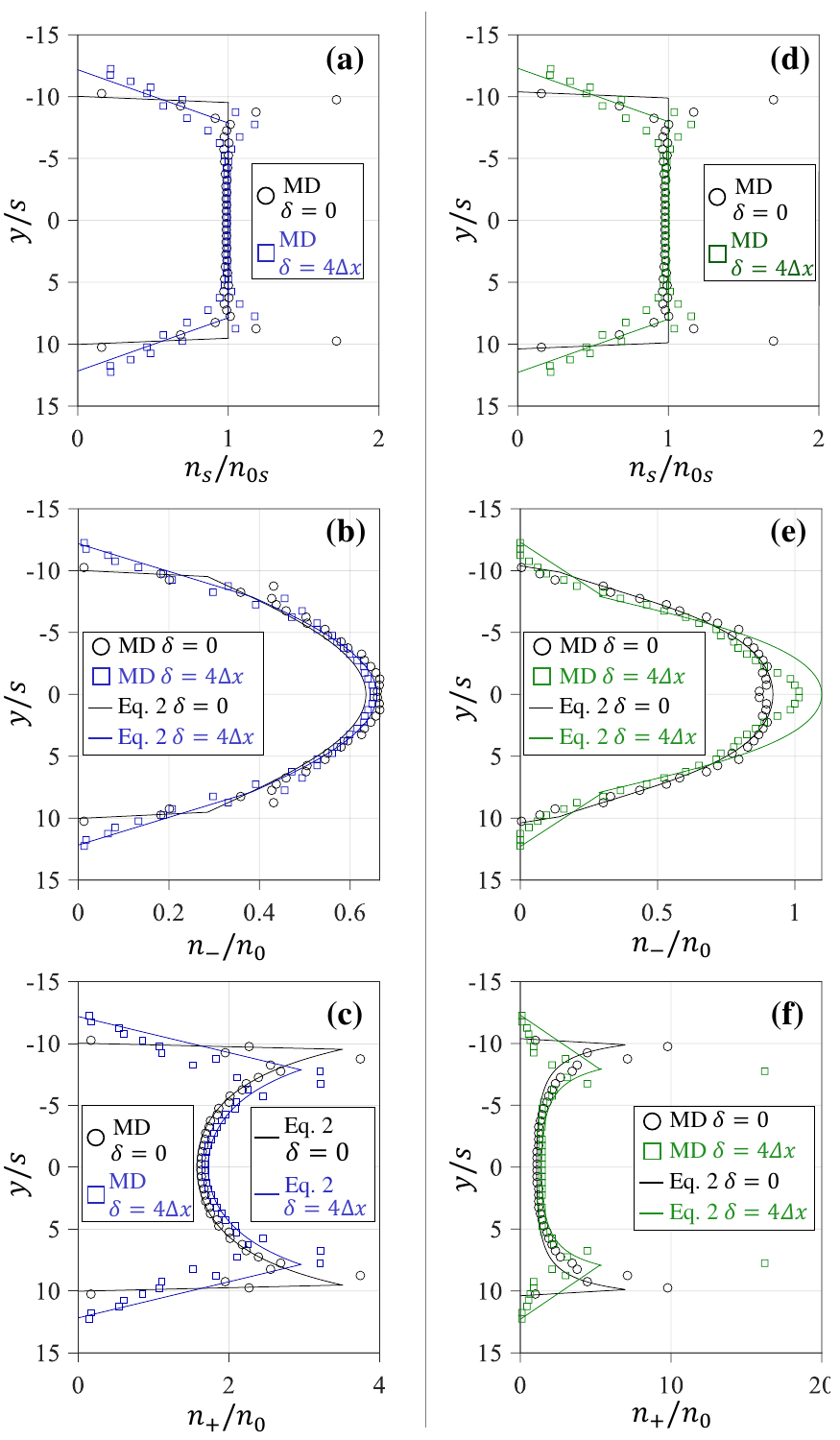}
 \caption{Streamwise-averaged solvent density $n_s$, anion density $n_-$, and cation density $n_+$ computed from MD simulations and predicted analytical using parameters reported in Table~\ref{tab:2}.
(a--c) Condition I with ``weak'' electrostatic interaction energy $C_e=1 k_B T$. 
(d--f) Condition II with ``strong'' electrostatic interaction energy $C_e=5 k_B T$. 
Solid lines in (a) \& (d) are piecewise linear fits to $n_s$ from MD results.
}
  \label{fig:3}
\end{figure}
\begin{figure}[h!]
\centering
\includegraphics[width=0.48\textwidth]{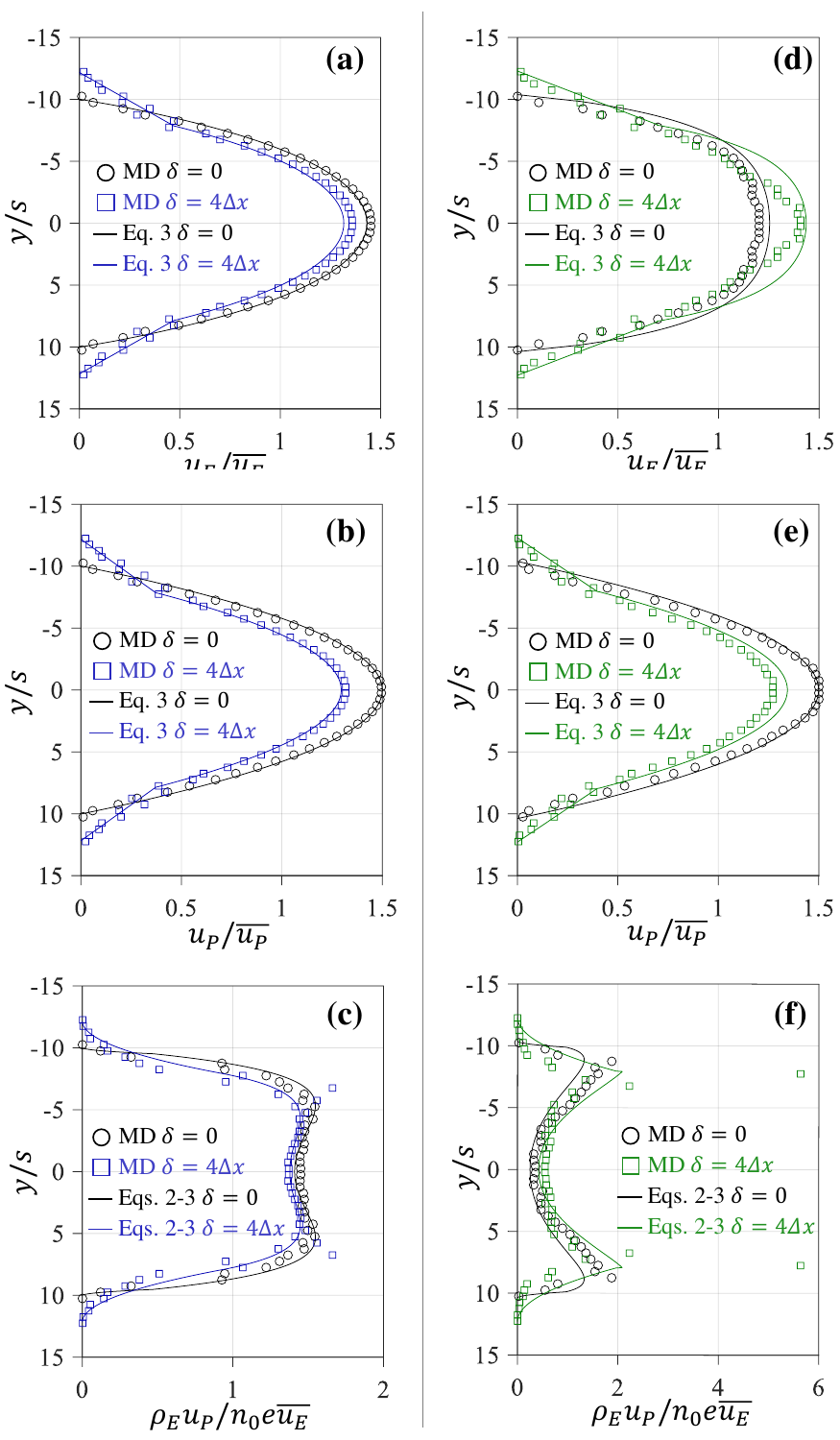}
\caption{Streamwise-averaged electroosmotic flow velocity $u_E$, pressure-driven flow velocity $u_P$, and streaming current flow $\rho_e u_P$ computed from MD simulations and predicted analytical using parameters reported in Table~\ref{tab:2}.
Velocity profiles are normalized by the mean values $\bar{u}_E=Q_E/WH$ and $\bar{u}_P=Q_P/WH$. 
(a--c) Condition I with ``weak'' electrostatic interaction energy $C_e=1 k_B T$. 
(d--f) Condition II with ``strong'' electrostatic interaction energy $C_e=5 k_B T$. 
}
\label{fig:4}
\end{figure}

The streamwise-averaged ion and solvent number density, flow velocity for electroosmotic and pressure-driven flows, and streaming current density is computed from the atomic position and velocities reported by the MD simulations, by spatial averaging along both the channel length and width plus additional time averaging over the last 9 ns of the simulations.  
The results reported in this section correspond to the cases of a flat channel with uniformly distributed surface charges and a ``rough'' channel with charges concentrated in the cuspids of the modeled surface features.
The results from MD simulations and the analytical expressions proposed in Sec.~\ref{sec:continuum} are compared in Figs.\ref{fig:3}--\ref{fig:4} for the case of a slit channel of average height $H\simeq 7.5$~nm, having atomically flat surfaces or surfaces with periodic feature of height $\delta\simeq 1.5$~nm and period $\ell\simeq 18$~nm.

Two different physical conditions are studied: 
Condition I having ``weak'' electrostatic interactions ($C_e=1 k_B T$) for which $\delta/\lambda_D\simeq 0.75$ and  $\ell/\lambda_D\simeq 9$; and     
Condition II having ``strong'' electrostatic interactions ($C_e=5 k_B T$) for which $\delta/\lambda_D\simeq 1.7$ and  $\ell/\lambda_D\simeq 20.2$.
For both conditions, the presence of the surface features modeling periodic surface roughness produces a nearly linear variation of the streamwise-averaged ion, net charge density, and fluid velocity profiles for $|y|\gtrsim y^*= H/2-\delta/2-s/2$, which is consistent with the continuum analysis for $\ell\ll\lambda_D$ presented in Sec.~\ref{sec:continuum}.
In all studied cases, MD simulations report significantly different behavior in the so-called ``inner'' region where $|y|\ge y^*$ and ``outer'' region where $|y|\le y^*$, which can be approximately accounted for by the proposed matched asymptotic expressions in Eqs.~\ref{eq:psi}--\ref{eq:flow}.
It is worth noticing, that the lack of dipole moments in the modeled liquid medium ensures that the electric permittivity in our MD simulations has the same constant value in the solution bulk and within the Stern layer.

\begin{table*}[]
\small
  \caption{Model parameters for analytical expressions in Eqs.~\ref{eq:psi}--\ref{eq:flow} and zeta potential $\zeta_{MD}$ computed from MD simulations and $\zeta$ determined by Eq.~\ref{eq:zeta_equality} for the two different channel geometry and modeled conditions with characteristic electrostatic energy $C_e$.
The zero-potential ion density $n_0$ and centerline potential $\psi(0)$ are obtained by solving numerically Eq.~\ref{eq:charges} for the number of charges $N_+$ and $N_{-}$, and the roughness height $\delta$ employed in the MD simulations.} 
   \begin{tabular*}{1\textwidth}{@{\extracolsep{\fill}}lcccccc}
    \hline
    Geometry & $C_e=e^2/(4\pi\varepsilon s)$ & $\delta/\lambda$ & $n_0$ [mM] & $\psi(0)$ [mV] & $\zeta_{MD}$ [mV]  & $\zeta$ [mV] \\
    \hline
    flat walls & 1 $k_BT$& 0 & 46.47 & -11.34 & -16.25 & -16.57\\
    rough walls & 1 $k_BT$ & 0.75 & 46.47 & -11.98 & -11.64 & -11.62\\
    flat walls & 5 $k_BT$ & 0 & 36.8 & -2.56 & -44.54 & -45.3\\
    rough walls & 5 $k_BT$ & 1.7 & 40.66 & -3.25 & -38.2 & -41.11\\
    \hline
  \end{tabular*}
\label{tab:2}
\end{table*}

The streamwise-averaged solvent number density $n_s(y)$ computed from MD simulations shows nearly identical profiles for Condition I and II (cf. Fig.~\ref{fig:3}a\&d) in the cases of ``flat'' and ``rough'' channel walls with the expected constant value $n_{0s}$ in the fluid bulk and spatial oscillations near the walls due to the presence of hydration layers.
For the modeled conditions, hydration layers with a quasi-crystalline structure are present within a distance between 1 to 3 molecular diameters from the walls, as it can be readily observed for the case of channels with flat walls.
Despite the spatial oscillations induced by the hydration layers, the streamwise-averaged solvent density profile for the case of channels with rough walls can be approximately described by a linear decay within the inner region where $|y|\gtrsim y^*$.
Nearly linear profiles computed from MD simulations in the inner region are also observed for the streamwise-averaged anion and cation density for both Condition I (cf. Fig.~\ref{fig:3}b-c) and Condition II (cf. Fig.~\ref{fig:3}e-f), albeit with a different slope steepness that scales inversely with the Debye length for each modeled condition.

Analytical expressions for the streamwise-averaged anion and cation density (Eq.~\ref{eq:ions}) reported in Fig.~\ref{fig:3} and flow velocity profiles (Eq.~\ref{eq:flow}) reported in Fig.~\ref{fig:4} employ the matching coordinate $y^*=H/2-s/2$ for the flat channel case and $y^*=H/2-\delta/2-s/2$ for the channel with surface features, with the model parameters reported in Table~\ref{tab:2}.
The zero-potential ion density $n_0$ and the centerline electrostatic potential $\psi(0)$ in each case is found numerically by satisfying Eq.~\ref{eq:charges} for $N_+=64$ and $N_-=16$ within a 1\% relative error.
For the case of streamwise-averaged density profiles, close agreement is observed in all cases for Condition I (Fig.~\ref{fig:3}a-c), while the agreement deteriorates for Condition II Fig.~\ref{fig:3}e-f) where very strong spatial oscillations are observed for the cation density profile in the inner region.
The lesser agreement between analytical predictions and MD simulations for Condition II where $C_e>k_B T$ is be attributed to the fact that the simple linear decay in the ion number density modeled by Eq.~\ref{eq:ions} near the walls does not accurately correspond to the complex functional form of the local electrostatic potential and solvation energy.

The streamwise-averaged velocity profiles for voltage-driven (electroosmotic) flow $u_E(y)$ and pressure-driven flow $u_P(y)$ from MD simulations are reported in Fig.~\ref{fig:4} for the studied flat and rough channel walls and the two conditions with different characteristic electrostatic energy $C_e$ (See Table~\ref{tab:2}). 
Similarly to the case of the number density profiles, a nearly linear velocity profile is observed in an inner region where $|y|\gtrsim y^*=H/2-\delta/2-s/2$.
Analytical predictions from Eq.~\ref{eq:flow} employ the matching velocity $u^*$ found by satisfying Eq.~\ref{eq:Q} with the electroosmotic flow rate $Q_E=Q(0,\Delta V)$ and pressure-driven flow rate $Q_P=Q(\Delta p,0)$ computed from MD simulations.
For physical consistency, the analytical predictions must also satisfy the Onsager reciprocal relation $Q_E/\Delta V=I_s/\Delta p$ between the electroosmotic flow rate and the streaming current, which is indeed satisfied by results from all our MD simulation within a small relative error $<5$\%. 
Analytical expressions via Eqs.~\ref{eq:ions}--\ref{eq:flow} for the streamwise-averaged current flux $\rho_e u_P$ are therefore expected to agree closely with results from MD simulations. 
As observed in Fig.~\ref{fig:4}a-c, analytical predictions for $u_E$, $u_P$, and $\rho_e u_P$ give close agreement with MD simulation results for Condition I with the weak electrostatic interactions (cf. Table~\ref{tab:2}) for which $C_e=1 k_BT$ and $\delta/\lambda_D\lesssim 1$.  
For Condition II where $C_e=5 k_BT$ and $\delta/\lambda_D>1$, close agreement is observed for the pressure-driven flow velocity $u_P$ (Fig.~\ref{fig:4}d), while there is a reasonable but lesser agreement between analytical expressions and MD simulations results for the streamwise-averaged electroosmotic flow velocity $u_E$ and streaming current flux $\rho_e$ (Fig.~\ref{fig:4}e-f).  

The zeta potential defined by the electroosmotic flow rate $Q_E$ via Eq.~\ref{eq:zeta_equality} is reported in Table~\ref{tab:2} for the case of flat ($\delta=0$) and rough ($\delta=4s$) walls under the studied two different conditions with ``weak'' and ``strong'' electrostatic interactions, for which  $\lambda_D=1.98$ nm and  $\lambda_D=0.9$ nm, respectively. 
For Condition I, the zeta potential $\zeta$ determined analytically via Eq.~\ref{eq:flow} and Eq.~\ref{eq:Q} agrees closely with the zeta potential $\zeta_{MD}$ computed in MD simulations.
For Condition II, analytical estimates for the zeta potential for the rough channel case fall within a 7\% relative error from the values reported by MD simulations.
We thus find that for the less optimal Condition II where $C_e = 5 k_B T$ and a channel with known roughness height $\delta =1.7 \lambda_D$, the employed analytical predictions approximately satisifes the reciprocal Onsager relation within a 10\% relative error.
For all the studied conditions, the modeled surface roughness has the effect of significantly reducing the zeta potential magnitude.

\section{Conclusions}
This work proposes a simple and compact analytical description for electrokinetic flows that can be employed to predict the zeta potential for the case of slit nanochannels with surface structure or roughness with a characteristic height $\delta\lesssim \lambda_D$ and period $\ell> \lambda_D$.
The proposed analytical predictions for the streamwise-averaged ion density, flow velocity, and streaming current flux were favorably compared against results from MD simulations for slit nanochannels with the same average height $H\simeq 7.5$ nm and having charged walls that are atomically flat ($\delta=0$) or have sinusoidal features of height $\delta\simeq 1.5$ nm.
In our MD simulations the number of charges at the walls is fixed, which is equivalent to neglecting the dependence of surface charge with the local surface potential, ion concentration, and solution pH.
The simulated conditions with ``weak'' and ``large'' values of the characteristic electrostatic energy ($C_e=$~1 and 5 $k_BT$) correspond to a liquid solution with relative permittivity 160 and 30, respectively. 
Furthermore, the medium permittivity is constant in our MD simulations due to the use of nonpolar monatomic molecules to model the liquid solvent. 
For all simulated cases no hydrodynamic slip is observed at the solid-liquid interface, which is due to the high affinity modeled between the solvent and solid atoms.
The lack of hydrodynamic slip at the walls in MD simulations ensured the fulfillment of the reciprocal Onsager relation between voltage-driven flow rate and pressure-driven current for all the studied channel geometries.

The proposed analytical model consists of matched asymptotic expressions for streamwise-averaged variables in an inner and outer region delimited by the plane $|y|=y^*$, where the matching coordinate 
$y^*$ is defined by the characteristic roughness height $\delta$ and the liquid solvent molecular diameter $s$.
The analytical results reported in Figs.~\ref{fig:3}--\ref{fig:4} and summarized in Table~\ref{tab:2}, are obtained from knowing the electroosmotic flow rate $Q_E$ and pressure-driven flow rate $Q_P$, and employing $y^*=H/2-\delta/2-s/2$ defined by the known roughness height $\delta$ and considering the presence of a Stern layer of thickness $s/2$.
Alternatively, in the case that the roughness height $\delta$ is not known, one can estimate it from Eqs.~\ref{eq:charges}~\ref{eq:Is} when the streaming current $I_s$ is known (e.g., from numerical simulation or experimental measurement).
Moreover, in the case that the streaming current is not known, one can compute or measure the pressure-driven flow rate $Q$ in order to estimate the effective flow rate profile $u_P$ using Eq.~\ref{eq:flow} and Eq.~\ref{eq:Q} and from there determine analytically the streaming current by using Eq.~\ref{eq:Is}. 
In this case, the effective thickness $\delta$ can be determine by invoking the reciprocal relation $Q_e/\Delta E=I_s/\Delta p$ that is satisfy for arbitrary channel geometry when there is no-hydrodynamic slip at the wall.
For example, using the latter procedure and the pressure-driven flow rate from MD simulations gives an effective roughness value that is reasonably close (i.e., within 7\% error) to that previously know from the simulated channel geometry.

The proposed analytical expressions do not consider the variation of the medium permittivity within the Stern layer nor spatial oscillations of the ion density that are induced by the presence of hydration layers near the solid-liquid interface.
Moreover, the approximate analytical expressions are valid for cases where the roughness period is large than the Debye length.
Nevertheless, the proposed analytical estimates can help rationalize zeta potential measurements for conditions commonly found in experimental studies of electeokinetic flow in slit channel with surfaces with nanoscale roughness of natural or synthetic origin.

\section*{Conflicts of interest}
The authors declare no conflicts of interest of any kind.

\section*{Acknowledgements}
A.A. was supported by the Office of Naval Research under award N00014-16-1-3178. 
S.H. was supported by the National Science Foundation under award CBET-1605809. 
T.S. and C.C. acknowledge support from award 75039 from the New York State Energy Research and Development Authority (NYSERDA) and award 76890 from the New York State Department of Economic Development (DED), which were provided as matching funds to the Center of Mesoscale Transport Properties, an Energy Frontier Research Center from the U.S. Department of Energy, Office of Science, Basic Energy Sciences, under award DE-SC0012673.  NYSERDA has not reviewed the information contained herein, and the opinions expressed do not necessarily reflect those of NYSERDA, or the State of New York.  Any opinions, findings, conclusions, or recommendations expressed are those of the author(s) and do not necessarily reflect the views of the DED.
%

%%%REFERENCES%%%
%\bibliography{bibliography} 

\begin{thebibliography}{10}
\expandafter\ifx\csname url\endcsname\relax
  \def\url#1{\texttt{#1}}\fi
\expandafter\ifx\csname urlprefix\endcsname\relax\def\urlprefix{URL }\fi
\expandafter\ifx\csname href\endcsname\relax
  \def\href#1#2{#2} \def\path#1{#1}\fi

\bibitem{levich1962}
V.~G. Levich, Physicochemical hydrodynamics, Vol. 689, Prentice-Hall Englewood
  Cliffs, NJ, 1962.

\bibitem{probstein2005}
R.~F. Probstein, Physicochemical hydrodynamics: an introduction, John Wiley \&
  Sons, 2005.

\bibitem{quirino1998}
J.~P. Quirino, S.~Terabe, Exceeding 5000-fold concentration of dilute analytes
  in micellar electrokinetic chromatography, Science 282~(5388) (1998)
  465--468.

\bibitem{hompesch2005analysis}
R.~W. Hompesch, C.~D. Garc{\'\i}a, D.~J. Weiss, J.~M. Vivanco, C.~S. Henry,
  Analysis of natural flavonoids by microchip-micellar electrokinetic
  chromatography with pulsed amperometric detection, Analyst 130~(5) (2005)
  694--700.

\bibitem{silva2013micellar}
M.~Silva, Micellar electrokinetic chromatography: a review of methodological
  and instrumental innovations focusing on practical aspects, Electrophoresis
  34~(1) (2013) 141--158.

\bibitem{hayward2000}
R.~Hayward, D.~Saville, I.~Aksay, Electrophoretic assembly of colloidal
  crystals with optically tunable micropatterns, Nature 404~(6773) (2000)
  56--59.

\bibitem{velev2006c}
O.~D. Velev, K.~H. Bhatt, On-chip micromanipulation and assembly of colloidal
  particles by electric fields, Soft Matter 2~(9) (2006) 738--750.

\bibitem{siavoshi2011size}
S.~Siavoshi, C.~Yilmaz, S.~Somu, T.~Musacchio, J.~R. Upponi, V.~P. Torchilin,
  A.~Busnaina, Size-selective template-assisted electrophoretic assembly of
  nanoparticles for biosensing applications, Langmuir 27~(11) (2011)
  7301--7306.

\bibitem{lee2002designing}
H.-J. Lee, F.~Sarfert, H.~Strathmann, S.-H. Moon, Designing of an
  electrodialysis desalination plant, Desalination 142~(3) (2002) 267--286.

\bibitem{sadrzadeh2008}
M.~Sadrzadeh, T.~Mohammadi, Sea water desalination using electrodialysis,
  Desalination 221~(1) (2008) 440--447.

\bibitem{almarzooqi2014application}
F.~A. AlMarzooqi, A.~A. Al~Ghaferi, I.~Saadat, N.~Hilal, Application of
  capacitive deionisation in water desalination: a review, Desalination 342
  (2014) 3--15.

\bibitem{elimelech1994measuring}
M.~Elimelech, W.~H. Chen, J.~J. Waypa, Measuring the zeta (electrokinetic)
  potential of reverse osmosis membranes by a streaming potential analyzer,
  Desalination 95~(3) (1994) 269--286.

\bibitem{sze2003zeta}
A.~Sze, D.~Erickson, L.~Ren, D.~Li, Zeta-potential measurement using the
  smoluchowski equation and the slope of the current--time relationship in
  electroosmotic flow, Journal of colloid and interface science 261~(2) (2003)
  402--410.

\bibitem{delgado2007}
{\'A}.~V. Delgado, F.~Gonz{\'a}lez-Caballero, R.~Hunter, L.~Koopal, J.~Lyklema,
  Measurement and interpretation of electrokinetic phenomena, Journal of
  colloid and interface science 309~(2) (2007) 194--224.

\bibitem{cummings2000conditions}
E.~B. Cummings, S.~Griffiths, R.~Nilson, P.~Paul, Conditions for similitude
  between the fluid velocity and electric field in electroosmotic flow,
  Analytical chemistry 72~(11) (2000) 2526--2532.

\bibitem{hu2003electrokinetic}
Y.~Hu, C.~Werner, D.~Li, Electrokinetic transport through rough microchannels,
  Anal. Chem. 75~(21) (2003) 5747--5758.

\bibitem{qiao2007effects}
R.~Qiao, Effects of molecular level surface roughness on electroosmotic flow,
  Microfluid. Nanofluid. 3~(1) (2007) 33--38.

\bibitem{yang2008numerical}
D.~Yang, Y.~Liu, Numerical simulation of electroosmotic flow in microchannels
  with sinusoidal roughness, Colloids. Surf. A 328~(1-3) (2008) 28--33.

\bibitem{liu2010molecular}
J.~Liu, M.~Wang, S.~Chen, M.~O. Robbins, Molecular simulations of
  electroosmotic flows in rough nanochannels, J. Comput. Phys. 229~(20) (2010)
  7834--7847.

\bibitem{messinger2010}
R.~Messinger, T.~Squires, Suppression of electro-osmotic flow by surface
  roughness, Physical review letters 105~(14) (2010) 144503.

\bibitem{yoshida2016a}
H.~Yoshida, T.~Kinjo, H.~Washizu, Analysis of electro-osmotic flow in a
  microchannel with undulated surfaces, Computers \& Fluids 124 (2016)
  237--245.

\bibitem{kirby2004}
B.~J. Kirby, E.~F. Hasselbrink, Zeta potential of microfluidic substrates: 1.
  theory, experimental techniques, and effects on separations, Electrophoresis
  25~(2) (2004) 187--202.

\bibitem{kirby2004b}
B.~J. Kirby, E.~F. Hasselbrink, Zeta potential of microfluidic substrates: 2.
  data for polymers, Electrophoresis 25~(2) (2004) 203--213.

\bibitem{aktar2019}
A.~Al~Hossain, A.~Checco, G.~Doerk, C.~Colosqui, Large-area nanostructured
  surfaces with tunable zeta potentials, Applied Materials Today 19 (2019)
  100553.

\bibitem{wenzel1936resistance}
R.~N. Wenzel, Resistance of solid surfaces to wetting by water, Ind. Eng. Chem.
  28~(8) (1936) 988--994.

\bibitem{ishino2004wetting}
C.~Ishino, K.~Okumura, D.~Qu{\'e}r{\'e}, Wetting transitions on rough surfaces,
  EPL 68~(3) (2004) 419.

\bibitem{healy1978}
T.~W. Healy, L.~R. White, Ionizable surface group models of aqueous interfaces,
  Advances in Colloid and Interface Science 9~(4) (1978) 303--345.

\bibitem{hiemstra1996intrinsic}
T.~Hiemstra, P.~Venema, W.~H. Van~Riemsdijk, Intrinsic proton affinity of
  reactive surface groups of metal (hydr) oxides: The bond valence principle,
  Journal of colloid and interface science 184~(2) (1996) 680--692.

\bibitem{borkovec1997origin}
M.~Borkovec, Origin of 1-p k and 2-p k models for ionizable water- solid
  interfaces, Langmuir 13~(10) (1997) 2608--2613.

\bibitem{behrens2001charge}
S.~H. Behrens, D.~G. Grier, The charge of glass and silica surfaces, The
  Journal of Chemical Physics 115~(14) (2001) 6716--6721.

\bibitem{chapman1913}
D.~L. Chapman, Li. a contribution to the theory of electrocapillarity, The
  London, Edinburgh, and Dublin Philosophical Magazine and Journal of Science
  25~(148) (1913) 475--481.

\bibitem{stern1924}
O.~Stern, Zur theorie der elektrolytischen doppelschicht, Zeitschrift f{\"u}r
  Elektrochemie und angewandte physikalische Chemie 30~(21-22) (1924) 508--516.

\bibitem{pashley1982hydration}
R.~Pashley, Hydration forces between mica surfaces in electrolyte solutions,
  Advances in Colloid and Interface Science 16~(1) (1982) 57--62.

\bibitem{israelachvili1983molecular}
J.~N. Israelachvili, R.~M. Pashley, Molecular layering of water at surfaces and
  origin of repulsive hydration forces, Nature 306~(5940) (1983) 249--250.

\bibitem{israelachvili2015}
J.~N. Israelachvili, Intermolecular and surface forces, Academic press, 2015.

\bibitem{behrens1999}
S.~H. Behrens, M.~Borkovec, Exact poisson-boltzmann solution for the
  interaction of dissimilar charge-regulating surfaces, Physical Review E
  60~(6) (1999) 7040.

\bibitem{mazur2013}
S.~R. De~Groot, P.~Mazur, Non-equilibrium thermodynamics, Courier Corporation,
  2013.

\bibitem{brunet2004generalized}
E.~Brunet, A.~Ajdari, Generalized onsager relations for electrokinetic effects
  in anisotropic and heterogeneous geometries, Physical Review E 69~(1) (2004)
  016306.

\bibitem{plimpton1993fast}
S.~Plimpton, Fast parallel algorithms for short-range molecular dynamics, Tech.
  rep., Sandia National Labs., Albuquerque, NM (United States) (1993).

\bibitem{plimpton2007lammps}
S.~Plimpton, P.~Crozier, A.~Thompson, Lammps-large-scale atomic/molecular
  massively parallel simulator, Sandia National Laboratories 18 (2007) 43.

\bibitem{lammps}
\href{https://lammps.sandia.gov}{[link]}.
\newline\urlprefix\url{https://lammps.sandia.gov}

\bibitem{hockney1988}
R.~W. Hockney, J.~W. Eastwood, Computer simulation using particles, crc Press,
  1988.

\bibitem{weinstein1953}
A.~Weinstein, Generalized axially symmetric potential theory, Bull. Am. Math.
  Soc. 59~(1) (1953) 20--38.

\bibitem{heinz1999charge}
W.~F. Heinz, J.~H. Hoh, Relative surface charge density mapping with the atomic
  force microscope, Biophys. J. 76~(1) (1999) 528--538.

\bibitem{gongadze2011adhesion}
E.~Gongadze, D.~Kabaso, S.~Bauer, T.~Slivnik, P.~Schmuki, U.~Van~Rienen,
  A.~Igli{\v{c}}, Adhesion of osteoblasts to a nanorough titanium implant
  surface, Int. J. Nanomed. 6 (2011) 1801.

\bibitem{drelich2011charge}
J.~Drelich, Y.~U. Wang, Charge heterogeneity of surfaces: Mapping and effects
  on surface forces, Adv. Colloid Interface Sci. 165~(2) (2011) 91--101.

\end{thebibliography}

\end{document}